\documentclass[12pt]{article}
\usepackage{makeidx}
\usepackage{multirow}
\usepackage{multicol}
\usepackage[dvipsnames,svgnames,table]{xcolor}
\usepackage{graphicx}
\usepackage{epstopdf}
\usepackage{ulem}
\usepackage{hyperref}
\usepackage{amsmath}
\usepackage{amssymb}
\usepackage{authblk}
\usepackage[paperwidth=595pt,paperheight=841pt,top=70pt,right=70pt,bottom=70pt,left=70pt]{geometry}

\makeatletter
	{\par\setlength{\parindent}{#3}
	\setlength{\leftmargin}{#1}       \setlength{\rightmargin}{#2}%
	\advance\linewidth -\leftmargin       \advance\linewidth -\rightmargin%
	\advance\@totalleftmargin\leftmargin  \@setpar{{\@@par}}%
	\parshape 1\@totalleftmargin \linewidth\ignorespaces}{\par}%
\makeatother 

\title{Farthest Streamline Sampling for the Uniform Distribution of Forearm Muscle Fiber Tracts from Diffusion Tensor Imaging}
\date{}

\author[1]{Yang Li}
\author[1,2]{Shihan Ma}
\author[1]{Jiamin Zhao}
\author[3]{Qing Li}
\author[1,4]{Xinjun Sheng}

\affil[1]{\fontsize{9pt}{12pt}\selectfont{State Key Laboratory of Mechanical System and Vibration, Shanghai Jiao Tong University, Shanghai, China}}
\affil[2]{Department of Bioengineering, Imperial College London, London, UK}
\affil[3]{MR Collaborations, Siemens Healthineers Digital Technology (Shanghai) Co., Ltd., Shanghai, China}
\affil[4]{Meta Robotics Institute, Shanghai Jiao Tong University, Shanghai, China}

\begin{document}
\maketitle
\noindent
\textbf{Background:} Diffusion tensor imaging (DTI) has been used to
characterize forearm muscle architecture. Since only uniform sampling is
performed for seed points rather than fiber tracts, the tracts may be unevenly
distributed in the muscle volume.

\noindent
\textbf{Purpose:} To reconstruct uniformly distributed fiber tracts in human
forearm by filtering the tracts from DTI.

\noindent
\textbf{Study Type:} Prospective.

\noindent
\textbf{Subjects:} 12 normal subjects.

\noindent
\textbf{Field Strength/Sequence:} 3T/T2-weighted turbo-spin-echo sequence with
simultaneous multi-slice technique and diffusion-weighted echo-planar sequence.

\noindent
\textbf{Assessment:} Farthest streamline sampling (FSS) was proposed for filtering and compared with two conventional methods, i.e., two-dimensional sampling and three-dimensional sampling. The uniform coverage performance of the methods was evaluated by streamline coverage (SC) and the coefficient of variation of streamline density (SDCV). Architectural parameters were calculated for 17 forearm muscles. Anatomical correctness was verified by 1. visually assessing the fiber orientation, 2. checking whether the architectural parameters were within physiological ranges, and 3. classifying the architectural types.

\noindent
\textbf{Statistical Tests:} We used a two-tailed paired \textit{t}-test to measure the
difference between samplings, and a one-sample \textit{t}-test to evaluate the ranges of
the architectural parameters and \textit{R}$^{2}$ for the classification of
architectural types.

\noindent
\textbf{Results:} FSS had the highest SC (0.93$\pm{}$0.04) and the lowest SDCV
(0.34$\pm{}$0.06) among the three methods (\textit{P}$<$0.05). FSS reduced the
sampling of long tracts (10\% reduction in fiber length, \textit{P}$<$0.05), and
the architectural parameters were within physiological ranges (two parameters
with \textit{P}$<$0.05). The fiber orientation of the tractography was visually
consistent with that of the cadaveric specimen. The architectural types of 16
muscles were correctly classified, except for the palmaris longus, which had a
linear arrangement of fiber endpoints (\textit{R}$^{2}$=0.95$\pm{}$0.02,
\textit{P}$<$0.001).

\noindent
\textbf{Data Conclusion:} FSS reconstructed more muscle regions and uniformly
distributed fiber tracts. The tracts were anatomically correct, indicating the
validity of fiber tracts.


\newpage
\section{Introduction}

\indent
Muscle architecture refers to “the macroscopic arrangement of muscle fibers” (1). The properties of muscle architecture are quantified by parameters such as fiber length and physiological cross-sectional area. These parameters are crucial in determining the muscle function in the musculoskeletal system, and have been widely applied to facilitate tendon transfer surgery (2), injury assessment (3), and biomechanical models (4). The traditional method for measuring muscle architecture involves manual dissection of cadaveric specimens (5). However, cadaveric data are mainly obtained from elder individuals (6), which makes the data less appropriate for young populations. In addition, changes in muscle morphology (4) lead to inaccurate extraction of in vivo information. \textit{In vivo} medical imaging techniques, including ultrasound imaging and diffusion tensor imaging (DTI), are alternative measurement methods. Lower-cost ultrasound imaging performs two-dimensional measurements in a narrow field of view, which limits its ability to adequately characterize muscle architecture (7). By contrast, although DTI comes with a higher cost, this technique allows for accurate visualization and three-dimensional measurement of numerous fibers from multiple muscles (8), and thus provides reconstruction of more realistic muscle fibers (7). Studies investigating muscle architecture using DTI mainly focus on large human muscles in the lower limbs (9,10). Only a few studies examined forearm muscles (11), and the architectural properties derived from DTI have not been widely reported. The primary reason is that forearm muscles have complex architecture due to their small volume and pennate morphology, making it challenging to measure muscle architecture.\\
\indent
Fiber tracts or streamlines generated by tractography are considered as the representation of fibers. It should be noted that the terms “tracts” and “streamlines” are interchangeable except in the case of density quantification (12). Streamline density does not represent the fiber density, and therefore it is preferred to use streamlines when describing density-dependent algorithms. The streamlines begin with seed points sampled at uniform or random spacing from the region of interest (ROI). The distribution of streamlines is estimated based on a hypothesis of uniform fiber density within biological tissues (13). However, since seed points are more likely to be sampled from long fibers and easily tracked voxels, longer and denser streamlines are produced, resulting in a streamline density bias. Unlike the sampling of seed points, filtering techniques samples the candidate streamlines to remove false-positive streamlines. Filtering (14) helps eliminate non-uniform density bias for white matter tractography. In muscle fiber tractography, filtering has not been used to enhance the streamline density, which may risk the inferences about muscle function due to the resulting biases in muscle properties. Moreover, it is imperative to assess the validity of fiber tracts, which is an important part of the validation of tractography. However, there is no recognized method for the assessment due to the lack of \textit{in vivo} actual values. Previous studies mainly performed empirical validation by establishing the repeatability of measurements (9,15), but the repeatability does not indicate validity.

\indent
Accurate architectural measurements are also subject to easily overlooked pitfalls that should be avoided. To constrain fiber tracts within the muscle volume, DTI data were first registered to anatomical images (15). Due to the complex architecture of forearm muscles, misregistration largely affects the performance of fiber tractography (12) and results in erroneous fiber orientation (9), indicating the demand for precise registration. The tract-based architectural properties of a muscle may not follow a normal distribution. Recent research has suggested using the median rather than the mean to report fiber length (8). Calculating mean architectural parameters without selection can be risky as the difference between the median and mean affects the measurements and may alter the inferences about muscle function.

\indent
This study aimed to reconstruct uniformly distributed fiber tracts for forearm muscles via a filtering method. Like in the white matter tractography, we hypothesized that less sampling of long and dense fiber tracts would reduce the streamline density bias. The hypothesis was tested by comparing the filtering method with conventional methods without filtering. Furthermore, the validity of fiber tracts was demonstrated by verifying the anatomical correctness.

\section{Materials and Methods}

\subsection{Study Participants}

\indent
The study was in accordance with the Declaration of Helsinki and approved by the local ethic committee (approval number E2021178C). All volunteers provided written informed consent. Prior to the measurements, participants were instructed to abstain from high-intensity exercise for at least 24 hours. MRI data were collected from the right forearm of 12 healthy subjects, consisting of 9 males and 3 females, with an average age of 25.3$\pm{}$2.2 years, height of 1.74$\pm{}$0.05 m, body mass of 66.9$\pm{}$7.2 kg, body mass index of 22.0$\pm{}$2.2 kgm$^{-2}$.

\subsection{MRI Protocol}

\indent
The measurements were conducted using a 3T scanner (MAGNETOM Prisma, Siemens Healthcare, Erlangen, Germany) equipped with an 18-channel flex body coil. Participants were positioned supine with their right arm placed neutrally beside the body, and an acrylic block was used to slightly lift the hand and prevent compression of the forearm. Two acquisitions were performed. T2-weighted (T2w) images for anatomical reference using simultaneous multi-slice (SMS) technique, with imaging parameters including turbo-spin-echo sequence, repetition time (TR)/echo time (TE): 3500 ms/39 ms, acquisition matrix: 448x448, transverse direction, reconstructed voxel size: 0.27x0.27x3 mm$^{3}$, SMS factor: 2, and scan time: 216 s. For accurate DTI of skeletal muscle (16), echo-planar sequence was employed with TR/TE: 5500 ms/69 ms, acquisition matrix: 80x160, sagittal direction, reconstructed voxel size: 1.6x1.6x3 mm$^{3}$, b value: 400 s/mm$^{2}$, diffusion directions: 12, and scan time: 292 s. The signal-to-noise ratio (SNR) was calculated as the ratio of the mean muscle signal in the ROI to the standard deviation of the noise from the background (17). The mean SNR for all subjects was 42$\pm{}$10 for T2w images and 25$\pm{}$2 for B0 images of DTI data, indicating sufficient quality (16) of the acquired images.

\subsection{Preprocessing}

\indent
Preprocessing of the DTI data was to correct eddy current, movement and sensitivity distortions based on B0 field maps using FSL eddy 6.0 (18). Rician noise was removed from the DTI data through a local PCA filter (19). Given that using global affine transformation is difficult to achieve precise registration, DTI data were registered to anatomical images using a non-rigid method, i.e., piecewise registration (20). First, The two sequences were aligned along the axial direction based on the surface markers (21) around the wrist and elbow (Figure 4 legend). Then, both sequences were divided into sections, each with about 15 slices. The local affine transformation was conducted separately for each section (Supplementary Material 1). Piecewise registration allows the correction of distortions in local regions, particularly in both ends, compared with global affine transformation (Supplementary Material 2). To evaluate the registration, mutual information and normalized mutual information (22) were calculated across all subjects. The preprocessing was followed by B-matrix rotation (23).

\subsection{Segmentation and Tractography}

\indent
According to the T2w images, 17 types of forearm muscles were manually segmented using ImageJ (24) to generate ROI-based muscle masks. The masks were examined by a researcher with 18 years of clinical experience in MRI. The masks and DTI data were resampled to an isotropic voxel size of 1 mm$^{3}$(10). Diffusion tensors were calculated using DSI Studio (25). Tractography was performed using deterministic fiber tracking. A curvature constraint was added to reduce overshoot errors and high curvature (26). Due to the constraint, streamlines propagated bidirectionally in 0.1 mm step and terminated when the angle between two adjacent steps exceeded 10$^\circ{}$. Additionally, the propagation was terminated if the fractional anisotropy was less than 0.1 or the muscle mask was exceeded (15). Streamlines less than 10 mm in length were discarded (27). The streamlines were fitted to third-order polynomial curves to allow for physiologically reasonable curvature (28). Endpoints of the streamlines were linearly extrapolated to the surface of the muscle mask to ensure attachment to the aponeurosis and robust measurement of fiber length (29). Streamlines with an extrapolated length greater than 30\% of the original length were excluded. A large sample of 3000 streamlines was reconstructed per muscle, which is considered sufficient to represent muscle architecture accurately (8). 

\subsection{Farthest Streamline Sampling}

\indent
To correct prospective streamline density bias, farthest streamline sampling (FSS) was proposed. The general process of FSS can be summarized in two steps. Seeding was performed at uniform spacing within the muscle mask for tractography to generate 10,000 candidate streamlines, and then the streamlines were filtered to 3,000 based on a distance index. The algorithm is described in detail below. First, all streamlines generated from tractography are uniformly resampled to $m$ points ($m$ is 12 for the optimal trade-off between streamline resolution and memory reduction (30)), e.g., streamline ${\textbf{\textsl{U}}}$ and ${\textbf{\textsl{V}}}$:

\begin{equation}
\textbf{\textsl{U}}=[{{\textbf{\textsl{u}}}_{1}},{{\textbf{\textsl{u}}}_{2}},...,{{\textbf{\textsl{u}}}_{m}}],\textbf{\textsl{ V}}=[{{\textbf{\textsl{v}}}_{1}},{{\textbf{\textsl{v}}}_{2}},...,{{\textbf{\textsl{v}}}_{m}}]
\label{eq:1}
\end{equation}

\noindent
Each element is a 3D coordinate. The flipped version of ${\textbf{\textsl{V}}}$ is  $\textbf{\textsl{ V}}^{F}=[{{\textbf{\textsl{v}}}_{m}},{{\textbf{\textsl{v}}}_{m-1}},...,{{\textbf{\textsl{v}}}_{1}}]$. The minimum average direct-flip (MDF) distance is calculated according to

\begin{equation}
{{d}_{\operatorname{direct}}}(\textbf{\textsl{U}},\textbf{\textsl{V}})=d(\textbf{\textsl{U}},\textbf{\textsl{V}})=\frac{1}{m}\sum\limits_{i=1}^{m}{\left| {{\textbf{\textsl{u}}}_{i}}-{{\textbf{\textsl{v}}}_{i}} \right|}
\label{eq:2}
\end{equation}

\begin{equation}
{{d}_{\operatorname{flipped}}}(\textbf{\textsl{U}},\textbf{\textsl{V}})=d(\textbf{\textsl{U}},{{\textbf{\textsl{V}}}^{\text{F}}})=d({{\textbf{\textsl{U}}}^{\text{F}}},\textbf{\textsl{V}})
\label{eq:3}
\end{equation}

\begin{equation}
\operatorname{MDF}(\textbf{\textsl{U}},\textbf{\textsl{V}})=\min ({{d}_{\operatorname{direct}}}(\textbf{\textsl{U}},\textbf{\textsl{V}}),\ {{d}_{\operatorname{flipped}}}(\textbf{\textsl{U}},\textbf{\textsl{V}}))
\label{eq:4}
\end{equation}

\noindent
Here, $\left| {{\textbf{\textsl{u}}}_{i}}-{{\textbf{\textsl{v}}}_{i}} \right|$  denotes the Euclidean distance. For sampled streamline set $\textbf{\textsl{S}}$ and remaining streamline set $\textbf{\textsl{R}}$:

\begin{equation}
\textbf{\textsl{S}}=[{{\textbf{\textsl{s}}}_{1}},{{\textbf{\textsl{s}}}_{2}},...,{{\textbf{\textsl{s}}}_{m}}],\textbf{\textsl{ R}}=[{{\textbf{\textsl{r}}}_{1}},{{\textbf{\textsl{r}}}_{2}},...,{{\textbf{\textsl{r}}}_{n}}]
\label{eq:5}
\end{equation}

\noindent
each element is a streamline, and the distance between  $\textbf{\textsl{S}}$ and next sampling streamline $\textbf{\textsl{r}}$  is defined as

\begin{equation}
D(\textbf{\textsl{r}},\textbf{\textsl{S}})=\underset{1\le i\le n}{\mathop{\max }}\,(\underset{1\le j\le m}{\mathop{\min }}\,(\underset{\begin{smallmatrix} 
 1\le i\le n \\ 
 1\le j\le m 
\end{smallmatrix}}{\mathop{\operatorname{MDF}}}\,({{\textbf{\textsl{r}}}_{i}},{{\textbf{\textsl{s}}}_{j}}))
\label{eq:6}
\end{equation}

\noindent
MDF is a distance index on the streamlined space, distinguishing short streamlines from the long (30). The principle of generating a uniform distribution by FSS can be explained geometrically: the next sampling streamline is located at the center of the largest empty region within a three-dimensional space. By reducing the sampling of long and dense streamlines, a specified number of streamlines with spatial uniformity is ensured. Two tractography metrics were used to evaluate the uniform coverage performance (14) for the sampling. Streamline coverage (SC) (31) indicates the capacity of streamlines to cover the muscle and was calculated as the ratio of the number of voxels crossed by streamlines to those of the muscle mask. Streamline density (SD) (32) is the mean number of streamlines crossing a voxel. The coefficient of variation of SD (SDCV), defined as the standard deviation/mean, indicates the non-uniformity of the distribution of streamlines. FSS was compared with two seed sampling strategies without filtering: uniform sampling based on two-dimensional slices (2DS) and based on three-dimensional muscle volumes (3DS). For 2DS, five slices were selected as the ROI (11), evenly spaced along the longitudinal axis.

\subsection{Architectural Properties}

The mean values of six architectural parameters were calculated for each muscle among the 12 subjects. The parameters include muscle volume (MV), fiber length (FL), muscle length (ML), FL/ML ratio, pennation angle (PA), and physiological cross-sectional area (PCSA). MV was estimated based on muscle masks from T2w images (33). FL was calculated as the sum of the distances between consecutive points of streamline (34). PA was measured as the angle of fiber tract relative to muscle’s line of action (35). For non-pennate muscles, the line of action was approximated by the average direction of all fiber tracts, while for pennate muscles, the least square-fitted line of fiber endpoints was used. When the goodness of fit $\textit{R}^{2}$ was greater than 0.9 (35), the fitted line was adopted due to the linear arrangement of the attachment for pennate muscles. ML was defined as the distance between the most proximal and distal points along the line of action (36). FL/ML ratio is an intrinsic property (not varying with muscle size) (1) and was calculated by dividing FL by ML. PCSA is defined as (1)

\begin{equation}
\text{PCSA}=\text{MV}*\text{cos}(\text{PA})/\text{FL}
\label{eq:7}
\end{equation}

\indent
To verify anatomical correctness, reconstructed tracts from tractography were compared with photographs from a human cadaver (11). The fiber orientation of two individual muscles, flexor carpi radialis and supinator, were inspected. The architectural types of forearm muscles are known to be pennate or non-pennate muscles (35,37). The arrangement of fiber endpoints was assessed by the classification of architectural types based on the relationship between the line of action and the architectural types. Accurate classification implies anatomically correct fiber endpoints.

\subsection{Functional Groups}

Functional groups of forearm muscles are defined in references (1,38) (Figure 6 legend). FL is proportional to the muscle excursion, while PCSA is proportional to the maximum isometric force (1). To compare the relative excursions and forces for muscles in the groups, a scatter graph of the force-generating capacity was plotted. Volume fractions of the functional groups and PCSA fractions of the muscles in each group are intrinsic properties and may be consistently distributed among subjects. Both fractions were compared with previous data of adult subjects (38).

\subsection{Statistical Analysis}

To determine the statistical appropriateness of the mean and median for tract-based architectural properties, including FL and PA, the Shapiro-Wilk test and calculation of skewness were conducted on individual muscles per subject. The other data were examined with the Shapiro-Wilk test. Since the normal distribution is not violated for the other data, parametric tests were used. Data were presented as mean$\pm{}$standard deviation unless otherwise stated. Differences in muscle architectural properties between sampling methods were calculated and depicted using Bland-Altman plots. The two-tailed paired \textit{t}-test was used to measure the significance of the differences of: 1. registration metrics and tractography metrics for the entire forearm region, 2. architectural parameters for the forearm muscles between samplings, and 3. fractions between studies. Typically, for muscles in human upper limbs, PA is less than 30$^\circ{}$, and FL/ML ratio ranges from about 0.2 to 0.6 (1). Thus, the one-sample \textit{t}-test evaluated whether the means of architectural parameters are within the physiological ranges. The test also compared the means of \textit{R}$^{2}$ and threshold 0.9 for classification of architectural types. The significance level was 0.05. Statistical tests were conducted using IBM SPSS version 25. 

\section{Results}

\subsection{Registration and Statistical Appropriateness}

Piecewise registration yielded significantly higher registration metrics than global affine transformation (mutual information: 2.92±0.22 and 2.77±0.21, normalized mutual information: 0.45±0.02 and 0.42±0.02, respectively). The Shapiro-Wilk test was conducted on the distribution of FL and PA within 203 forearm muscles for 12 subjects (one subject lacked palmaris longus (39)). Neither architectural property of all muscles demonstrated a normal distribution (Figure 1). Therefore, the use of median to represent the central tendency of FL and PA was supported statistically.

\subsection{Farthest Streamline Sampling}

For tractography metrics, FSS showed significantly higher streamline coverage (0.93±0.04) than both 2DS (0.83±0.08) and 3DS (0.88±0.07), with the lowest SDCV among the samplings (FSS: 0.34±0.06, 2DS: 0.44±0.08, 3DS: 0.39±0.06). These differences implied that FSS reconstructed more muscle regions and increased the uniformity of the distribution of streamlines. To further illustrate the effect of reducing fluctuations in SD, SD maps (Figure 2) were plotted for the three samplings and merged to create a movie showing the SD distribution within the forearm muscles (Supplementary Material 3). While 2DS tended to have less streamline coverage and yielded more black (low-density) regions, 3DS had moderate tractography metrics but produced more blue (high-density) regions. 

\noindent
Bland-Altman analysis showed that the architectural properties for 2DS and 3DS were similar, with no significant differences observed. The largest mean difference was -5\% in PA (\textit{P}=0.256, 95\% confidence interval (CI): [-15, 5]). The two seed sampling methods and FSS differed significantly in the architectural properties, except for PA (FSS and 3DS, \textit{P}=0.758, 95\% CI: [-10.2, 13.5]). Considering that there is no significant difference between 2DS and 3DS, we report the mean value of the difference between FSS and the seed sampling methods. FL was 10\% shorter and ML was 6\% longer, resulting in a 14\% reduction in FL/ML ratio. Although PA was increased by 5\%, which may reduce PCSA, PCSA rose by 8\%. This means that potential error in FL was translated into the difference in FL/ML ratio and PCSA, and the cosine of PA (Equation (7)) had little effect on PCSA. The relative differences observed were consistent for each muscle in the subjects. Figure 3 shows the difference pattern with Bland-Altman plots. 

\subsection{Architectural Properties}

The \textit{in vivo} architectural parameters of forearm muscles were determined (Supplementary Material 4). The values of PA and FL/ML ratio were within the physiological range. Anatomical images were overlaid on muscle masks to examine the segmentation (Figure 4(b)). Fiber tractography was shown in different coding forms (Figure 4(c–g)). Figure 5 shows the comparison of the tractography and cadaveric photographs of supinator and flexor carpi radialis. Supinator, which curved around the upper portion of the radius, originated from the lateral humeral epicondyle and was inserted into the proximal radius (Figure 5(c) black dotted lines). The proximal fibers tended to be horizontal (magnified area in Figure 5(c)C), while the distal fibers were axial (magnified area in Figure 5(c)B) (40). The proximal attachments of flexor carpi radialis included the radial common flexor tendon and the ulnar aponeurosis (left black dotted lines in Figure 5(d)) and flexor carpi radialis inserted in the distal tendon (right lines). The fibers originating from the proximal attachments had ulnar-to-radial oblique orientation (magnified area in Figure 5(d)A), while the fibers attached to the distal tendon were axially oriented (magnified area in Figure 5(d)B) (41). The reconstructed muscles were in favorable correspondence with the cadaveric specimens. Movies of the comparison are available in Supplementary Materials 5 and 6. Most of the classifications of architectural types were correct (Supplementary Material 7), except for palmaris longus, which should be non-pennate but had a linear arrangement of fiber endpoints (\textit{R}$^{2}$=0.95$\pm{}$0.02, \textit{P}$<$0.001).

\subsection{Functional Groups}

The distribution of force-generating capacity of the functional groups is shown in Figure 6. Flexor digitorum profundus and flexor digitorum superficialis were distinguished from the other muscles, indicating that finger flexors had a significantly greater capacity for force and excursion than the other functional groups. Finger extensors had weak force-generating capacity, with extensor digitorum communis being the strongest in the group. Wrist flexors and wrist extensors displayed moderate properties, with a gradient seen mainly in force and excursion, respectively. Among the functional group of elbow and forearm muscles, pronator teres was the strongest, while the other muscles had a weak capacity for excursion.

\indent
For the volume fractions of the functional groups (Figure 7(a)), the volume of finger flexors was the largest, while the volume of finger extensors was the smallest. The volume fractions of the other functional groups were roughly equal. Compared to the data from (38), no group had a difference greater than two percentage points. A consistent distribution of PCSA fractions was observed in each group among the subjects. The distribution from this study did not significantly differ from that in (38).

\section{Discussion}

In this study, we proposed a filtering method, farthest streamline sampling (FSS) to enhance the uniform distribution of fiber tracts for forearm muscles. The method reconstructed more muscle regions and reduced streamline density bias, compared to conventional methods without filtering. The reconstructed fiber tracts were anatomically correct, which informs the demonstration of the validity of fiber tracts. 

\indent
Based on the hypothesis that less sampling of long and dense fiber tracts reduces the streamline density bias, FSS algorithm was designed by using the distance index MDF (30) and the farthest sampling strategy. The index and the strategy enabled FSS to distinguish between streamlines of different lengths in a three-dimensional space with restricted distance. The intention of introducing the hypothesis was to model the uniform fiber density within biological tissues (13). Similar hypotheses appeared in white matter filtering (32,42) to correct the density bias and the studies sampled streamlines of different lengths on purpose or not. The hypothesis was demonstrated by the difference of fiber length and streamline density from the comparison of FSS with conventional methods without filtering. Consistent with the hypothesis, FSS reconstructed shorter fiber tracts, along with lower density fluctuations.

\indent
From the results of tractography metrics and architectural properties, FSS produced high coverage and uniform distribution of fiber tracts, resulting in stable density (42) and accurate architectural measurements. As a progressive algorithm, FSS found the streamline farthest from the current set at each step, which limited the minimum distance between streamlines and avoided local aggregation. This improvement in localization is reflected in the tractography metrics and can be observed in the SD maps (14). The essential difference in methodology between FSS and conventional methods was that FSS filtered the streamlines rather than just sampling the seed points. The methodology can explain the performance of conventional methods 2DS and 3DS. The non-uniformity of distribution for slice-based 2DS resulted from the small seeding ROI. Reconstructed fiber tracts covered small muscle regions that did not contain all morphological fibers (43). 3DS produced uniformly distributed seed points within the muscle volume, but high streamline density resulted from longer streamlines, indicating that the uniform distribution of seed points is not equivalent to that of fiber tracts. Moreover, according to (12), FSS may reduce the deviations in the position, morphology, and length of tracts, which were reflected in the differences of architectural properties between samplings. These differences were computationally correlated. The most critical correlation was that the reduction of FL led to a decrease of FL/ML ratio and an increase of PCSA. Overall, the enhancement to the distribution of fiber tracts by FSS ensures accurate architectural measurements.

\indent
To assess the validity of fiber tracts in the absence of a “gold standard”, examining the correlation between cadaveric and \textit{in vivo} muscle data is a preferable approach. Data from cadavers provide insight into underlying musculature trends, though the correlation may be inappropriate and subjective due to morphological changes in cadavers, such as reduced mass and muscle contraction (6). We assessed the validity of fiber tracts based on the anatomical correctness in three aspects. First, we compared the tractography with cadaveric photographs (11) of the individual muscles. The intuitive and high correlation is demonstrated by the arrangement from the origin to the insertion and the fiber orientation at both ends. Second, the parameters FL/ML ratio and PA were within the physiological range 0.2–0.6 and 0–30°, which achieved quantitative anatomical validation. In forearm muscles, the relatively longest fibers extend to about 60\% of the muscle length, and the angle between the most oblique fibers and the line of action does not exceed 30° (1). Third, muscle architectural types were successfully classified. The classification was based on the arrangement of fiber endpoints, which reflected the anatomical differences between pennate and non-pennate muscles. Most forearm muscles belong to pennate muscles. The linear arrangement of the endpoints for pennate muscles helps generate large muscle force (35). One explanation for the misclassification of PL is that the muscle is characterized by high morphological diversity (39). The PL is a slender, fusiform-shaped muscle, the belly of which is partially fused with the adjacent muscles. These morphological factors may allow the PL to have more parallel fibers and thus a linear endpoint arrangement.

\indent
Some evidence (38) suggests the consistency in the distribution of muscle force-generating properties within functional groups among subjects, despite significant inter-subject variability in muscle size. The force-generating properties of the functional groups in this study confirmed the previous findings and increased our confidence in the accurate measurements of forearm muscle architecture using DTI. First, the excursions and forces of the functional groups presented specialization (8), which facilitates the execution of muscle functions. According to the consistent inter-subject trends in the distribution of FL and PCSA, finger flexors were classified as power specialised (high FL and high PCSA) for gripping movements. Low values of finger extensors indicated that finger extension is a low-power movement. The medium values of wrist extensors and wrist flexors implied moderate excursions and forces of wrist movement. Second, the PCSA fractions for muscles in each functional group were found to be consistent among subjects (38). Since PCSA is the only architectural parameter proportional to the maximum isometric force (1), the consistency of fractions implied that the force distribution may remain stable within each group. Third, the volume fraction of each functional group was constant, and the volume of the flexors was approximately twice that of the extensors (38). As a parameter derived from muscle masks, the consistent volume fractions among subjects from different studies might indicate the accuracy of muscle segmentation. Furthermore, these consistent force-generating properties support the assignment of muscle architectural properties by similar distributions in musculoskeletal models (4).

\indent
For accurate architectural measurements, the pitfalls of tractography, such as misregistration and statistical appropriateness, were avoided. The large field of view of the MRI protocol allowed for data to be acquired in a single scan. However, the disadvantage of the large field was the increased geometric distortion, especially for areas away from the imaging center. The mixed distortion arose from the eccentric position of the ends and the chemical artifacts from the bones (44), causing misregistration that cannot be well resolved by a normal preprocessing and global affine transformation. Therefore, piecewise registration (20) was used and proved to correct the local distortion included at the ends. For tract-based architectural properties, the FL and PA of forearm muscles were statistically supported as the median in this paper, to avoid differences from the mean affecting the measurements. This statistical appropriateness is the same as that for lower limb muscles (8). 

\indent
This study has limitations. First, the number of subjects was small due to the high cost of data acquisition. The sample size was unbalanced across gender. Nevertheless, the large variability in muscle size (15\%-45\% coefficient of variation in muscle volume) and the anatomical correctness suggests a limited effect of the sample. Second, the number of candidate streamlines was fixed. The common filtering process is estimating candidate streamlines and selecting a subset (15). The number of candidate streamlines depends on several factors such as imaging parameters, SNR of the data and tractography (42), and the number can be further selected using tractography metrics to obtain a trade-off between time and performance. The characterization of the optimal candidate set represents an important study (15), which is beyond the scope of this paper. Third, to measure architectural properties at the maximum isometric force, fiber length could be normalized to the optimal length based on sarcomere length. However, it is challenging to measure \textit{in vivo} sarcomere length using imaging techniques, and only a few reports are available for the sarcomere length in forearm muscles from cadavers (5). To reduce the variation in sarcomere length, subjects were scanned in the neutral position. Lastly, the brachioradialis, which originates above the lateral humeral epicondyle (middle of the upper arm), was not reconstructed due to the inadequate coverage of the MRI protocol.

\indent
In conclusion, we proposed a novel filtering method for the uniform distribution of fiber tracts from DTI. The validity of fiber tracts was demonstrated through anatomical correctness. The filtering method highlights the importance of eliminating the streamline density bias in tractography to measure muscle architectural properties accurately. These properties can be used as input to biomechanical models and as a reference for clinical evaluation of muscle pathology.

\newpage
\section{References}

\hspace{1em}
1. Lieber RL, Friden J. Clinical significance of skeletal muscle architecture. Clinical Orthopaedics and Related Research 2001(383):140-151.

2. Livermore A, Tueting JL. Biomechanics of Tendon Transfers. Hand Clinics 2016;32(3):
291-+.

3. Ramasamy E, Avci O, Dorow B, et al. An Efficient Modelling-Simulation-Analysis Workflow to Investigate Stump-Socket Interaction Using Patient-Specific, Three-Dimensional, Continuum-Mechanical, Finite Element Residual Limb Models. Frontiers in Bioengineering and Biotechnology 2018;6.

4. Charles JP, Grant B, D'Aout K, Bates KT. Subject-specific muscle properties from diffusion tensor imaging significantly improve the accuracy of musculoskeletal models. Journal of Anatomy 2020;237(5):941-959.

5. Liu A-T, Liu B-L, Lu L-X, et al. Architectural properties of the neuromuscular compartments in selected forearm skeletal muscles. Journal of Anatomy 2014;225(1):12-18.

6. Martin ML, Travouillon KJ, Fleming PA, Warburton NM. Review of the methods used for calculating physiological cross-sectional area (PCSA) for ecological questions. Journal of Morphology 2020;281(7):778-789.

7. Bolsterlee B, Veeger HEJ, van der Helm FCT, Gandevia SC, Herbert RD. Comparison of measurements of medial gastrocnemius architectural parameters from ultrasound and diffusion tensor images. Journal of Biomechanics 2015;48(6):1133-1140.

8. Charles J, Kissane R, Hoehfurtner T, Bates KT. From fibre to function: are we accurately representing muscle architecture and performance? Biological Reviews 2022;97(4):1640-1676.

9. Oudeman J, Mazzoli V, Marra MA, et al. A novel diffusion-tensor MRI approach for skeletal muscle fascicle length measurements. Physiological reports 2016;4(24).

10. Bolsterlee B, Finni T, D'Souza A, Eguchi J, Clarke EC, Herbert RD. Three-dimensional architecture of the whole human soleus muscle in vivo. Peerj 2018;6.

11. Froeling M, Nederveen AJ, Heijtel DFR, et al. Diffusion-tensor MRI reveals the complex muscle architecture of the human forearm. Journal of Magnetic Resonance Imaging 2012;36(1):237-248.

12. Girard G, Whittingstall K, Deriche R, Descoteaux M. Towards quantitative connectivity analysis: reducing tractography biases. Neuroimage 2014;98:266-278.

13. Hwang K, Huan F, Kim DJ. Muscle fibre types of the lumbrical, interossei, flexor, and extensor muscles moving the index finger. Journal of Plastic Surgery and Hand Surgery 2013;47(4):268-272.

14. Daducci A, Dal Palu A, Lemkaddem A, Thiran J-P. COMMIT: Convex Optimization Modeling for Microstructure Informed Tractography. Ieee Transactions on Medical Imaging 2015;34(1):246-257.

15. Bolsterlee B, D'Souza A, Herbert RD. Reliability and robustness of muscle architecture measurements obtained using diffusion tensor imaging with anatomically constrained tractography. Journal of Biomechanics 2019;86:71-78.

16. Froeling M, Nederveen AJ, Nicolay K, Strijkers GJ. DTI of human skeletal muscle: the effects of diffusion encoding parameters, signal-to-noise ratio and T-2 on tensor indices and fiber tracts. Nmr in Biomedicine 2013;26(11):1339-1352.

17. Dietrich O, Raya JG, Reeder SB, Reiser MF, Schoenberg SO. Measurement of signal-to-noise ratios in MR images: Influence of multichannel coils, parallel imaging, and reconstruction filters. Journal of Magnetic Resonance Imaging 2007;26(2):375-385.

18. Andersson JLR, Sotiropoulos SN. An integrated approach to correction for off-resonance effects and subject movement in diffusion MR imaging. Neuroimage 2016;125:1063-1078.

19. Manjon JV, Coupe P, Concha L, Buades A, Collins DL, Robles M. Diffusion Weighted Image Denoising Using Overcomplete Local PCA. Plos One 2013;8(9).

20. Li Y, Ma S, Li Q, et al. Tracking forearm muscle fibers from diffusion MRI during dynamic contractions. In: Proceedings of the 31st Annual Meeting of ISMRM, London; 2022. (abstract 3167).

21. Izatt MT, Lees D, Mills S, Grant CA, Little JP. Determining a reliably visible and inexpensive surface fiducial marker for use in MRI: a research study in a busy Australian Radiology Department. Bmj Open 2019;9(8).

22. Klein S, Staring M, Pluim JPW. Evaluation of optimization methods for nonrigid medical image registration using mutual information and B-splines. Ieee Transactions on Image Processing 2007;16(12):2879-2890.

23. Leemans A, Jones DK. The B-Matrix Must Be Rotated When Correcting for Subject Motion in DTI Data. Magnetic Resonance in Medicine 2009;61(6):1336-1349.

24. Schneider CA, Rasband WS, Eliceiri KW. NIH Image to ImageJ: 25 years of image analysis. Nature Methods 2012;9(7):671-675.

25. Yeh F-C, Verstynen TD, Wang Y, Fernandez-Miranda JC, Tseng W-YI. Deterministic Diffusion Fiber Tracking Improved by Quantitative Anisotropy. Plos One 2013;8(11).

26. Tournier JD, Calamante F, Connelly A. MRtrix: Diffusion tractography in crossing fiber regions. International Journal of Imaging Systems and Technology 2012;22(1):53-66.

27. Li Z, Mogk JP, Lee D, Bibliowicz J, Agur AM. Development of an architecturally comprehensive database of forearm flexors and extensors from a single cadaveric specimen. Computer Methods in Biomechanics and Biomedical Engineering: Imaging \& Visualization 2015;3(1):3-12.

28. Damon BM, Heemskerk AM, Ding Z. Polynomial fitting of DT-MRI fiber tracts allows accurate estimation of muscle architectural parameters. Magnetic Resonance Imaging 2012;30(5):589-600.

29. Bolsterlee B, D'Souza A, Gandevia SC, Herbert RD. How does passive lengthening change the architecture of the human medial gastrocnemius muscle? Journal of Applied Physiology 2017;122(4):727-738.

30. Garyfallidis E, Brett M, Correia MM, Williams GB, Nimmo-Smith I. Quick Bundles, a method for tractography simplification. Frontiers in Neuroscience 2012;6.

31. Cote M-A, Girard G, Bore A, Garyfallidis E, Houde J-C, Descoteaux M. Tractometer: Towards validation of tractography pipelines. Medical Image Analysis 2013;17(7):844-857.

32. Calamante F, Tournier J-D, Heidemann RM, Anwander A, Jackson GD, Connelly A. Track density imaging (TDI): Validation of super resolution property. Neuroimage 2011;56(3):1259-1266.

33. Eng CM, Abrams GD, Smallwood LR, Lieber RL, Ward SR. Muscle geometry affects accuracy of forearm volume determination by magnetic resonance imaging (MRI). Journal of Biomechanics 2007;40(14):3261-3266.

34. Heemskerk AM, Sinha TK, Wilson KJ, Ding Z, Damon BM. Repeatability of DTI-based skeletal muscle fiber tracking. Nmr in Biomedicine 2010;23(3):294-303.

35. Lee D, Li Z, Sohail QZ, Jackson K, Fiume E, Agur A. A three-dimensional approach to pennation angle estimation for human skeletal muscle. Computer Methods in Biomechanics and Biomedical Engineering 2015;18(13):1474-1484.

36. Takahashi K, Shiotani H, Evangelidis PE, Sado N, Kawakami Y. Three-dimensional architecture of human medial gastrocnemius fascicles in vivo: Regional variation and its dependence on muscle size. Journal of Anatomy 2022;241(6):1324-1335.

37. Matulionis DH. A Functional Study of the Forearm Musculature of the Human and Macaca mulatta. Bios 1966;37(1):3-14.

38. Holzbaur KRS, Murray WM, Gold GE, Delp SL. Upper limb muscle volumes in adult subjects. Journal of Biomechanics 2007;40(4):742-749.

39. Olewnik L, Wysiadecki G, Polguj M, Podgorski M, Jezierski H, Topol M. Anatomical variations of the palmaris longus muscle including its relation to the median nerve - a proposal for a new classification. Bmc Musculoskeletal Disorders 2017;18.

40. Berton C, Wavreille G, Lecomte F, Miletic B, Kim H-J, Fontaine C. The supinator muscle: anatomical bases for deep branch of the radial nerve entrapment. Surgical and Radiologic Anatomy 2013;35(3):217-224.

41. Segal RL, Catlin PA, Krauss EW, Merick KA, Robilotto JB. Anatomical partitioning of tree human forearm muscles. Cells Tissues Organs 2002;170(2-3):183-197.

42. Pestilli F, Yeatman JD, Rokem A, Kay KN, Wandell BA. Evaluation and statistical inference for human connectomes. Nature Methods 2014;11(10):1058-1063.

43. Heemskerk AM, Damon BM. Diffusion tensor MRI assessment of skeletal muscle architecture. Current Medical Imaging Reviews 2007;3(3):152-160.

44. Sinha U, Csapo R, Malis V, Xue Y, Sinha S. Age-Related Differences in Diffusion Tensor Indices and Fiber Architecture in the Medial and Lateral Gastrocnemius. Journal of Magnetic Resonance Imaging 2015;41(4):941-953.

\newpage
\section{Figures}

\begin{figure}[h]
    \centering
    \includegraphics[width=1\linewidth]{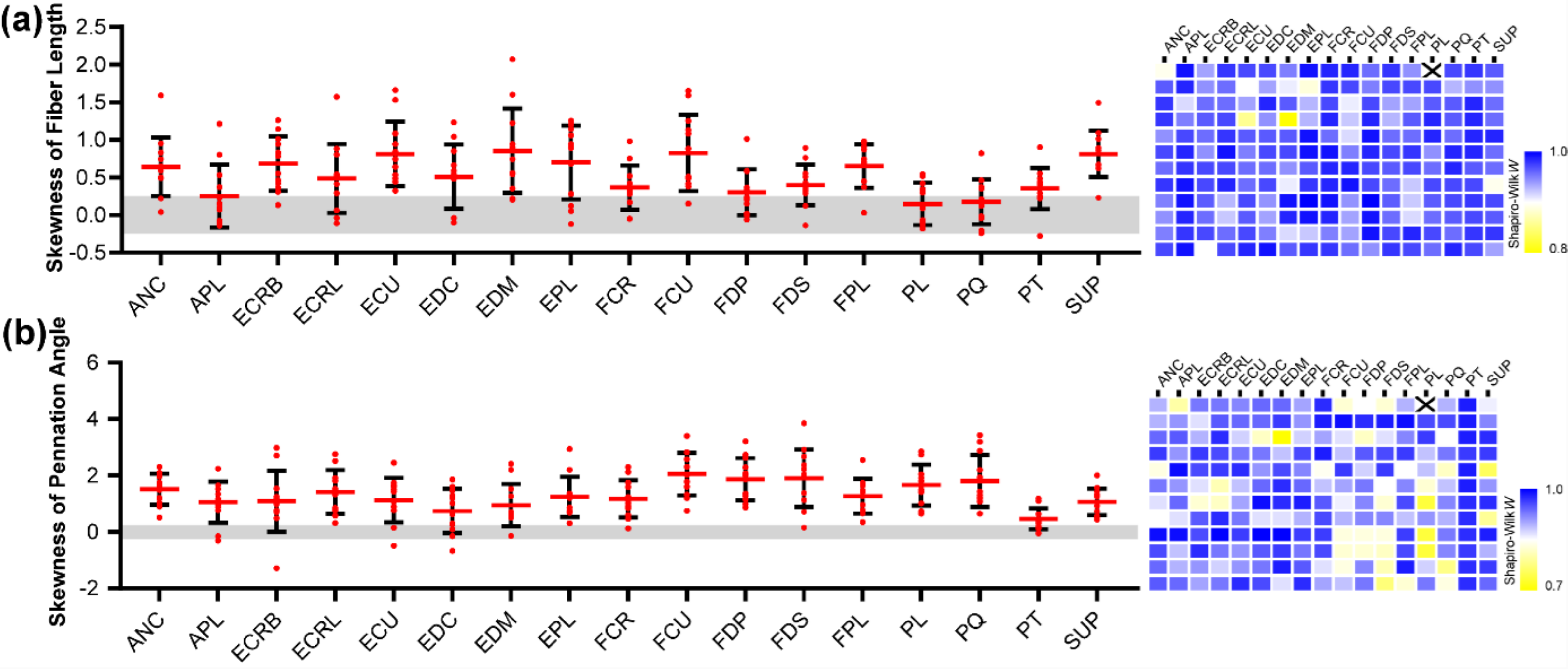}
    \label{fig:1}

\end{figure}

\noindent
\textbf{Figure 1.} Statistical appropriateness of (\textbf{a}) fiber length and (\textbf{b}) pennation angle. The \textit{W} value of Shapiro-Wilk test and the skewness for each muscle in subjects indicated that both architectural properties violated the normality test and were mainly distributed in a positive-skewed distribution. (Abbreviations: anconeus (ANC), abductor pollicis longus (APL), extensor carpi radialis brevis (ECRB), extensor carpi radialis longus (ECRL), extensor carpi ulnaris (ECU), extensor digitorum communis (EDC), extensor digiti minimi (EDM), extensor pollicis longus (EPL), flexor carpi radialis (FCR), flexor carpi ulnaris (FCU), flexor digitorum profundus (FDP), flexor digitorum superficialis (FDS), flexor pollicis longus (FPL), palmaris longus (PL), pronator quardratus (PQ), pronator teres (PT), supinator (SUP).)

\newpage
\begin{figure}[h]
    \centering
    \includegraphics[width=1\linewidth]{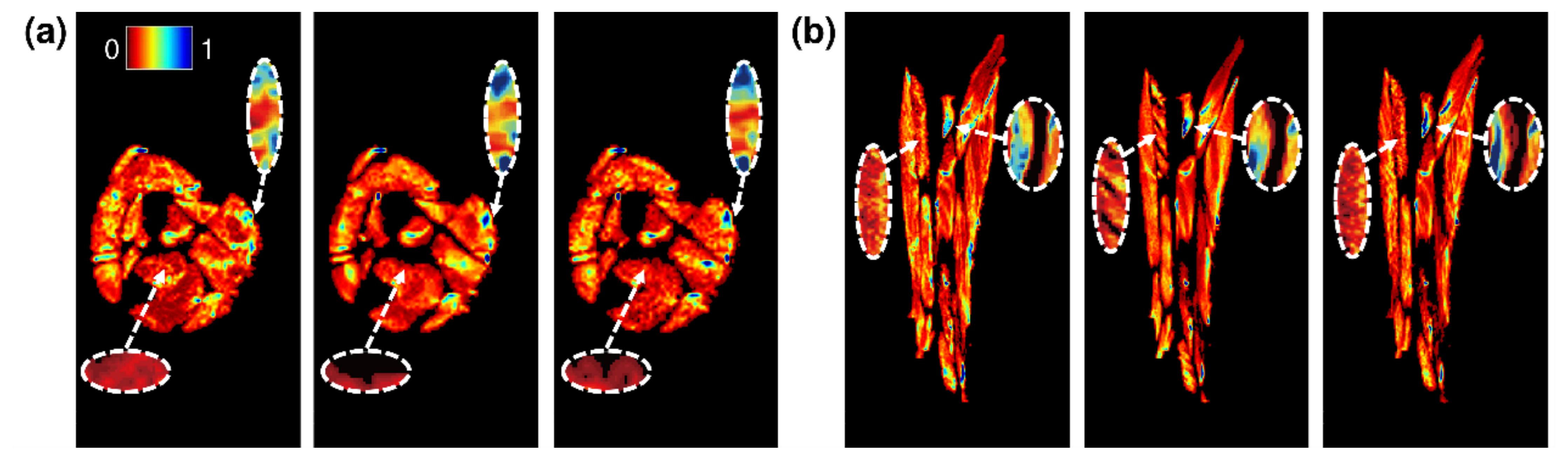}
    \label{fig:2}

\end{figure}

\noindent
\textbf{Figure 2.} Streamline density maps of the forearm muscles for the three samplings: FSS, 2DS and 3DS in (\textbf{a}) axial and (\textbf{b}) coronal views. Streamline density was normalized to the range [0, 1] in each muscle mask. 2DS yielded more black regions of low values while 3DS produced more blue regions of high values. Some local regions displayed density biases, which are shown in white dashed circles and have been scaled for visibility.

\newpage
\begin{figure}[h]
    \centering
    \includegraphics[width=1\linewidth]{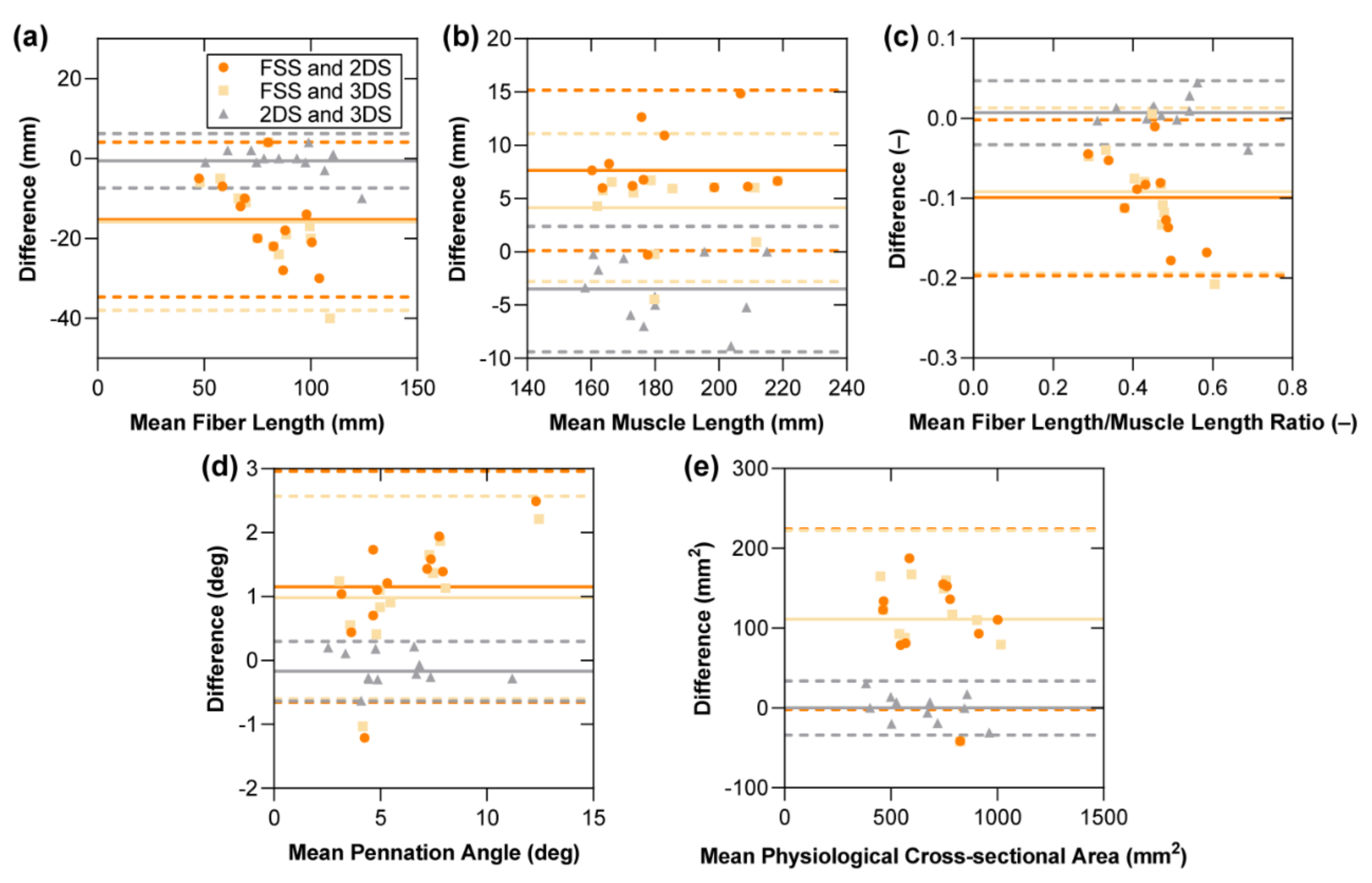}
    \label{fig:3}

\end{figure}

\noindent
\textbf{Figure 3.} Bland-Altman plots between FSS, 2DS, and 3DS for the flexor digitorum superficialis muscle, showing differences in (\textbf{a}) fiber length, (\textbf{b}) muscle length, (\textbf{c}) fiber length/muscle length ratio, (\textbf{d}) pennation angle, and (\textbf{e}) physiological cross-sectional area. The means (solid lines) and 95\% confidence intervals (dashed lines) show the distribution of differences. The differences between 2DS and 3DS were distributed around zero, while those between FDS and other samplings were on one side of zero.

\newpage
\begin{figure}[h]
    \centering
    \includegraphics[width=1\linewidth]{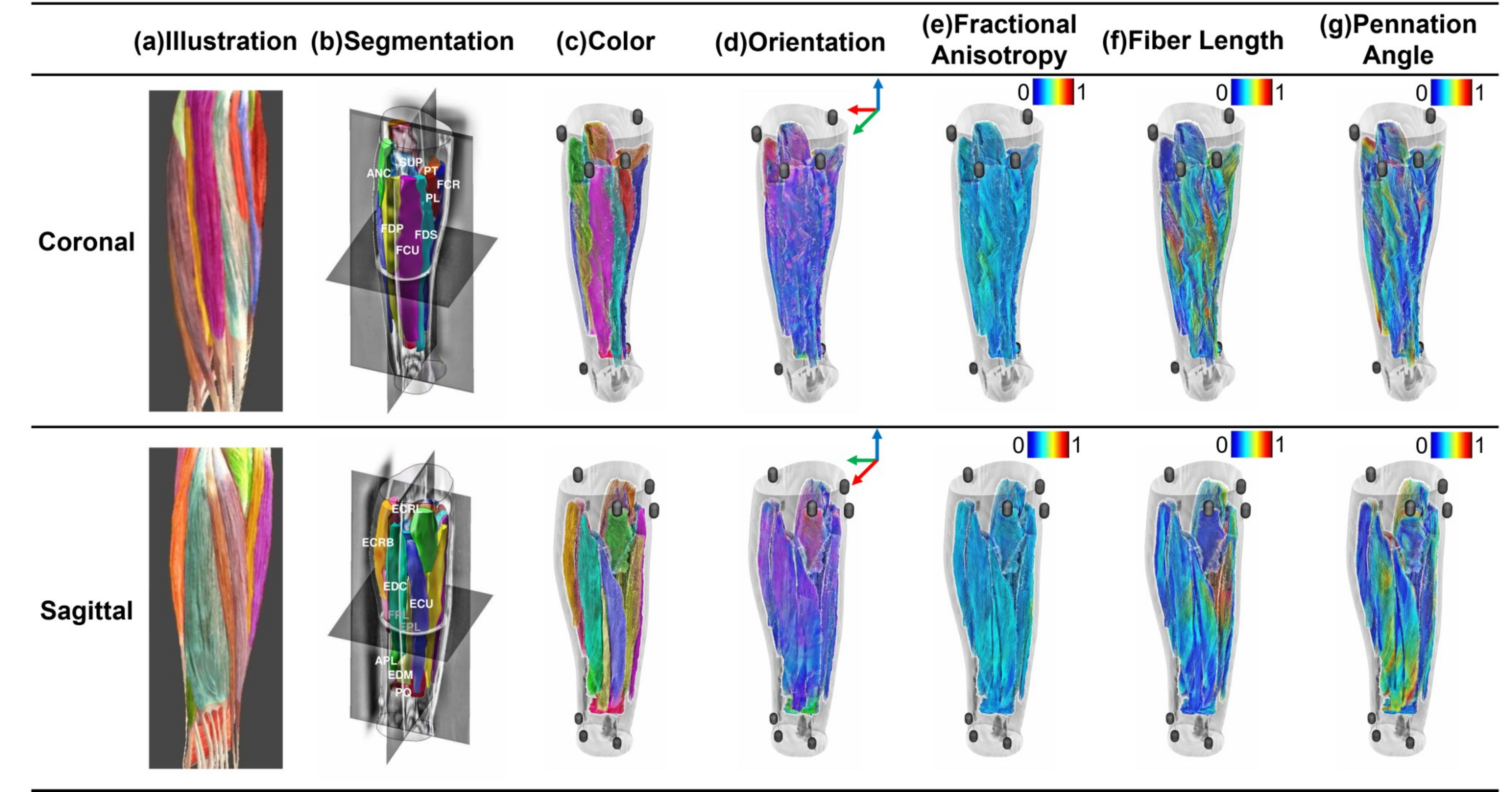}
    \label{fig:4}

\end{figure}

\noindent
\textbf{Figure 4.} Visualization results of the 17 forearm muscles in coronal and sagittal views, listed as (\textbf{a}) illustration, (\textbf{b}) segmentation, and fiber tractography in coding of (\textbf{c}) muscular color, (\textbf{d}) orientation, (\textbf{e}) fractional anisotropy, (\textbf{f}) fiber length and (\textbf{g}) pennation angle. Same muscular colors were added to (\textbf{a}), (\textbf{b}) and (\textbf{c}). Surface markers (vitamin D capsules (21)) were used to locate the anatomical position, including the styloid process of ulnar, styloid process of radius, Lister's tubercle, lateral humeral epicondyle, medial humeral epicondyle, olecranon, and fossa cubitalis in (\textbf{c}–\textbf{g}). (Figure 1. for muscle abbreviations in (\textbf{b}).)

\newpage
\begin{figure}[h]
    \centering
    \includegraphics[width=.8\linewidth]{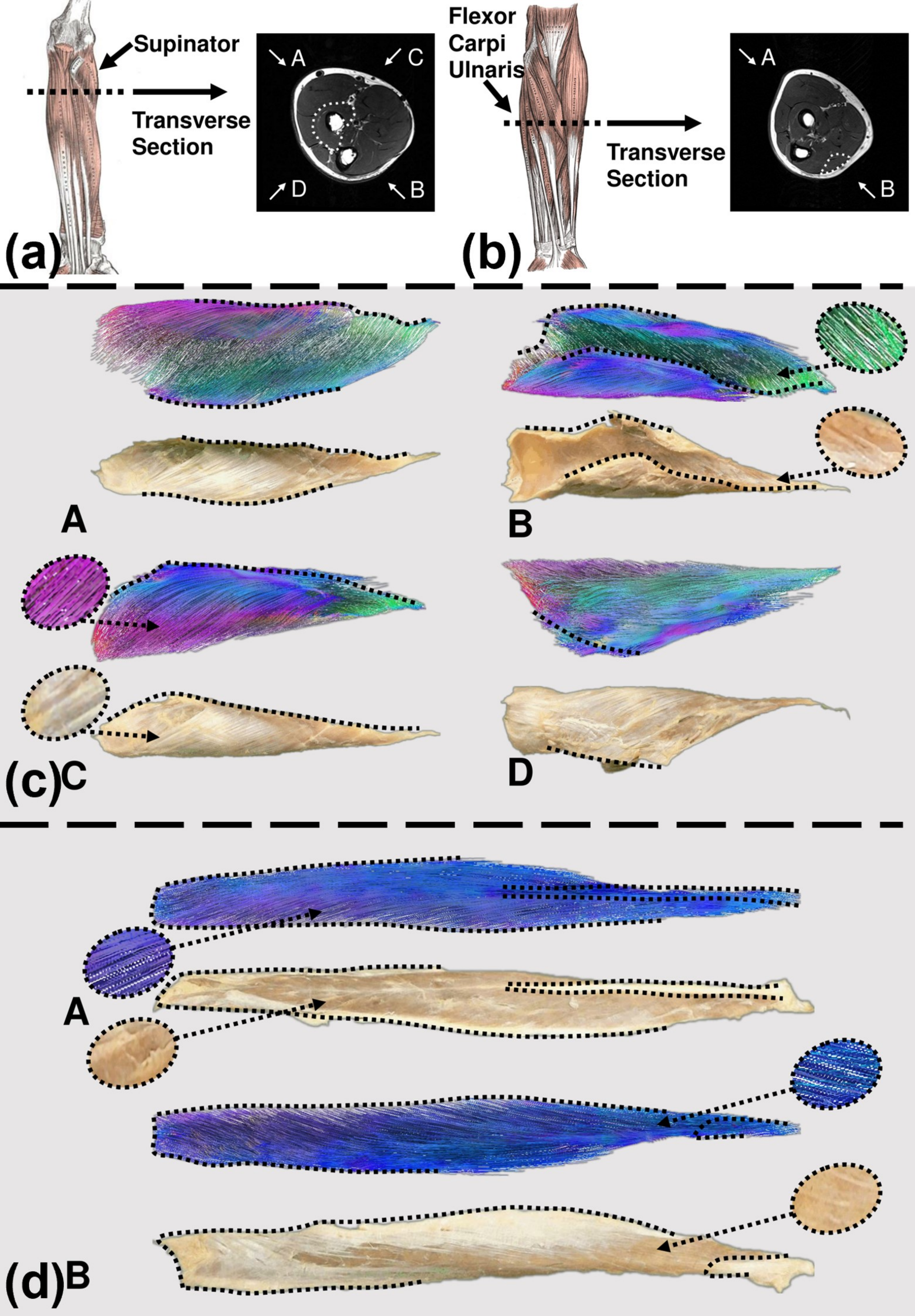}
    \label{fig:5}

\end{figure}

\noindent
\textbf{Figure 5.} Different perspectives A–D for observing (\textbf{a}) supinator (SUP) and (\textbf{b}) flexor carpi radialis (FCU) on T2-weighted images. The white dotted lines show contours of the muscles. Orientation-coded tractography and cadaveric photographs (11) (with permission) from the perspectives for (\textbf{c}) SUP and (\textbf{d}) FCU. The black dotted lines indicate the origin and insertion at (\textbf{c}) and attachments at (\textbf{d}).

\newpage
\begin{figure}[h]
    \centering
    \includegraphics[width=.8\linewidth]{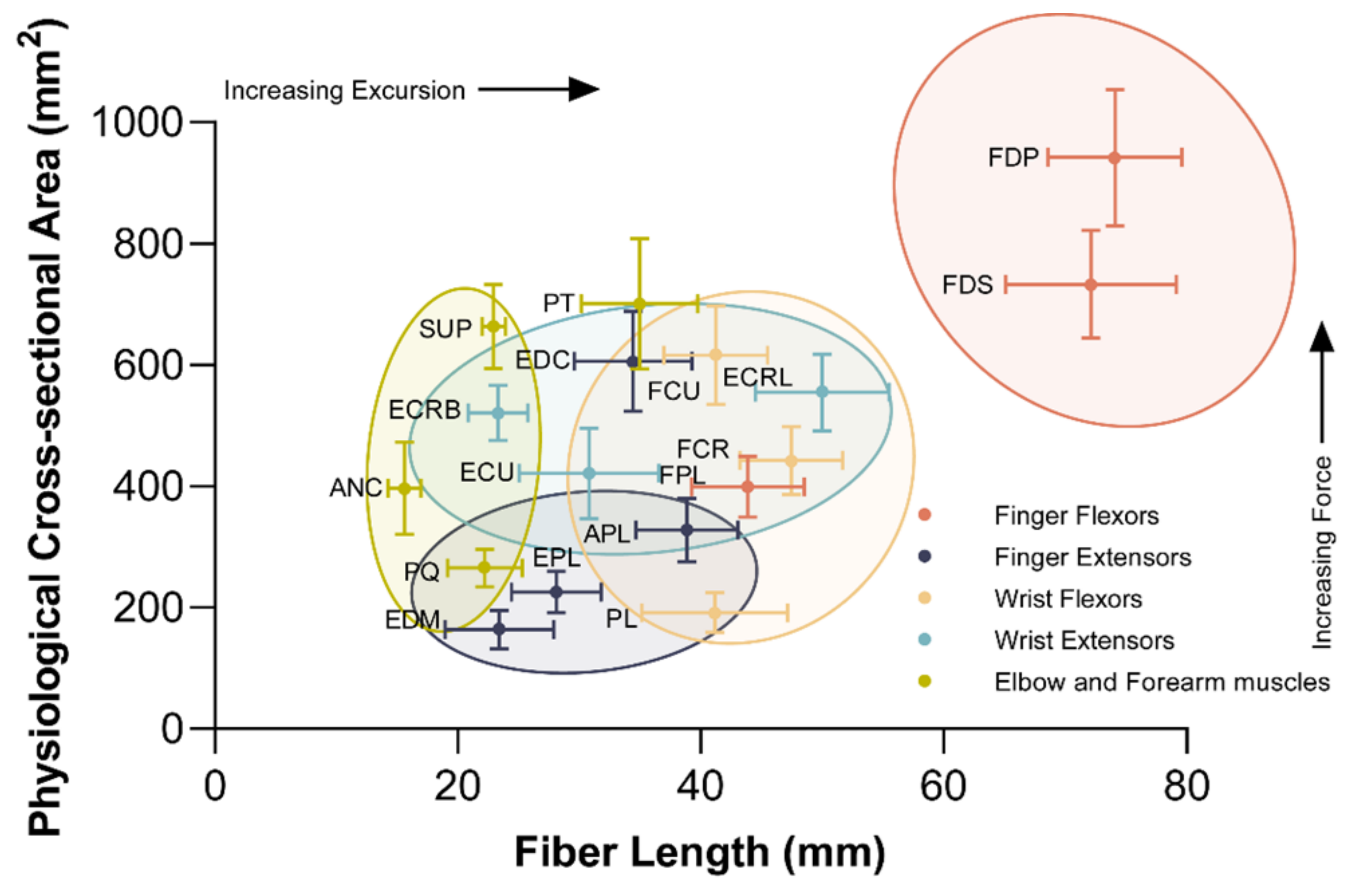}
    \label{fig:6}

\end{figure}

\noindent
\textbf{Figure 6.} A scatter graph of the fiber length and physiological cross-sectional area of forearm muscles. The muscles per functional group: finger flexors (FDP, FDS, FPL), finger extensors (APL, EDC, EDM, EPL), wrist flexors (FCR, FCU, PL), wrist extensors (ECRB, ECRL, ECU), elbow and forearm muscles (ANC, PQ, PT, SUP). Each group is covered with a 70\% confidence region. PT, EDC and FPL are excluded from the regions (similar practices in (1) for small region overlaps and clear comparisons). (Figure 1 for muscle abbreviations.)

\newpage
\begin{figure}[h]
    \centering
    \includegraphics[width=1\linewidth]{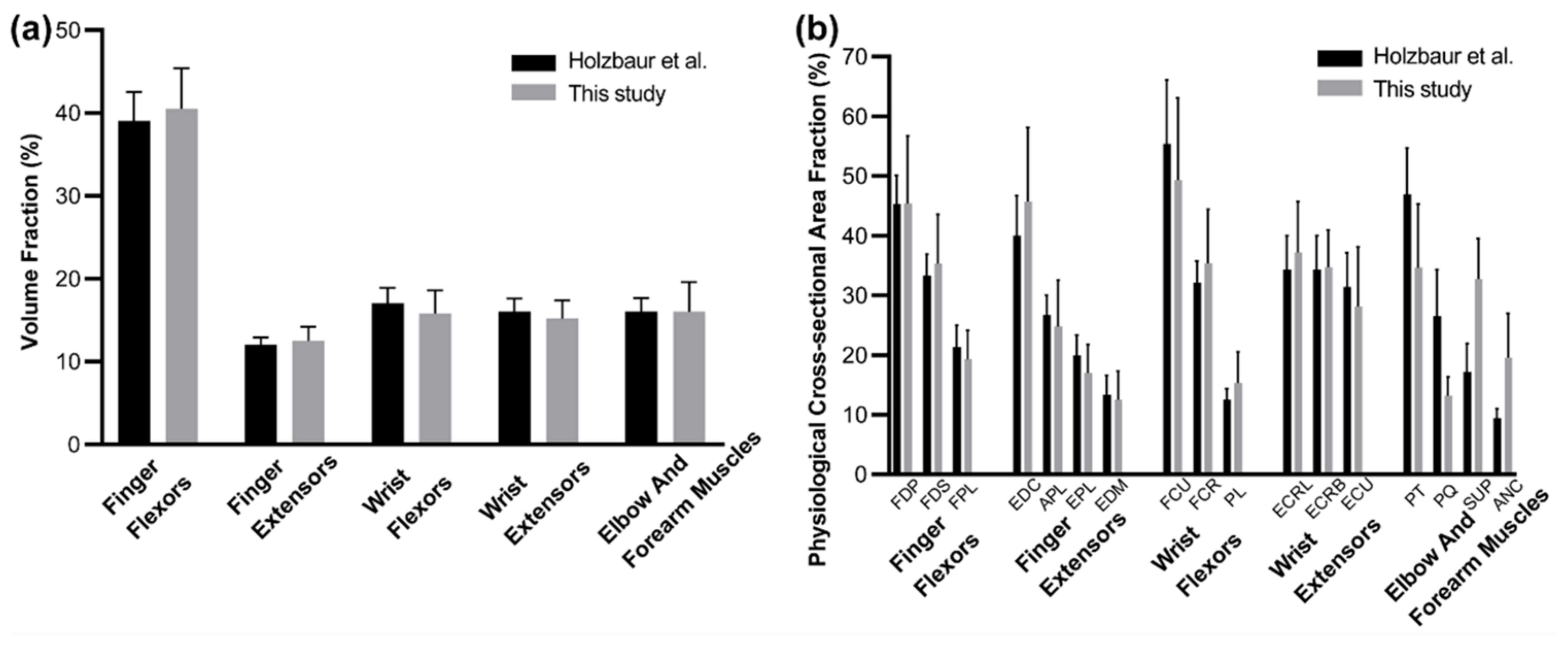}
    \label{fig:7}

\end{figure}

\noindent
\textbf{figure 7.} (\textbf{a}) Volume fractions of the functional groups and (\textbf{b}) physiological cross-sectional area fractions of the muscles in each group, compared to the data of adult subjects from Holzbaur et al. (38) (with permission). Error bars represent one standard deviation.

\end{document}